\documentclass[conference,twocolumn]{IEEEtran}

\usepackage{graphicx}
\usepackage{color}
\usepackage{placeins}
\usepackage{float}
\usepackage{tabularx,colortbl}
\usepackage{amssymb}
\usepackage{amsthm}
\usepackage{cite}
\usepackage{amsmath}

\usepackage{caption2}
\theoremstyle{plain}
\newtheorem{thm}{Theorem}

\theoremstyle{plain}
\newtheorem{rem}{Remark}
\newtheorem{cor}{Corollary}

\IEEEoverridecommandlockouts

\begin{document}

\title{Team-Optimal MMSE Combining for Cell-Free Massive MIMO Systems}

\author{Jiakang Zheng, Jiayi Zhang, and Bo Ai

\thanks{J. Zheng and J. Zhang are with the School of Electronics and Information Engineering, Beijing Jiaotong University, Beijing 100044, P. R. China (e-mail: \{20111047, jiayizhang\}@bjtu.edu.cn).}
\thanks{B. Ai is with the State Key Laboratory of Rail Traffic Control and Safety, Beijing Jiaotong University, Beijing 100044, China (e-mail: boai@bjtu.edu.cn).}
}

\maketitle
\vspace{-1cm}
\begin{abstract}
Cell-free (CF) massive multiple-input multiple-output (MIMO) systems are expected to implement advanced cooperative communication techniques to let geographically distributed access points jointly serve user equipments.
Building on the \emph{Team Theory}, we design the uplink team minimum mean-squared error (TMMSE) combining under limited data and flexible channel state information (CSI) sharing. Taking into account the effect of both channel estimation errors and pilot contamination, a minimum MSE problem is formulated to derive unidirectional TMMSE, centralized TMMSE and statistical TMMSE combining functions, where CF massive MIMO systems operate in unidirectional CSI, centralized CSI and statistical CSI sharing schemes, respectively. We then derive the uplink spectral efficiency (SE) of the considered system.
The results show that, compared to centralized TMMSE, the unidirectional TMMSE only needs nearly half the cost of CSI sharing burden with neglectable SE performance loss. Moreover, the performance gap between unidirectional and centralized TMMSE combining schemes can be effectively reduced by increasing the number of APs and antennas per AP.
\end{abstract}

\IEEEpeerreviewmaketitle

\section{Introduction}

Cell-free (CF) massive multiple-input multiple-output (MIMO), where a large number of access points (APs) are geographically distributed over a large area and coherently serve all user equipments (UEs) on the same time-frequency resource, has gained plenty of attention over the past years \cite{Ngo2017Cell,zhang2020prospective}.
Compared with traditional cellular networks, the main characteristics of CF massive MIMO systems is the operating regime with no cell boundaries and many more APs than UEs \cite{9355403,8768014}.
Moreover, CF massive MIMO can also reap all benefits of massive MIMO, such as favorable propagation and channel hardening by using multiple antennas at APs \cite{8379438}.
Besides, both large macro-diversity and high coverage probability can be achieved in CF massive MIMO systems, which makes it a promising wireless access technology for beyond fifth-generation (B5G) networks \cite{8972478}.

The notion of cooperation has been extensively studied in the context of CF massive MIMO systems as a tool to extend coverage, improve spectral efficiency, and manage inter-user interference \cite{9416909}.
For instance, based on different implementations of CF massive MIMO systems, authors in \cite{8845768} proposed different levels of cooperation among the APs, including centralized network and local processing network.
Besides, signal processing methods, e.g., uplink minimum mean-squared error (MMSE) and zero-forcing (ZF) combining, can be used individually at each AP to suppress inter-user interference \cite{5594708}.
Results in \cite{7869024} show that CF massive MIMO systems with centralized MMSE combining achieve nearly three times higher 90\%-likely SE than local MMSE combining. In addition, authors in \cite{9064545} observe both local partial MMSE (LP-MMSE) combining and centralized P-MMSE combining under the dynamic cooperation clustering, and demonstrate P-MMSE outperforms LP-MMSE. Moreover, closed-form expressions for the uplink SE of scalable CF massive MIMO systems under partial ZF combining are derived in \cite{9529197}. However, the uplink combiner design of CF massive MIMO systems with practical cooperation regimes, limited data and flexible channel state information (CSI) sharing scheme is still a big challenge.

In order to solve this challenge, we resort to the well-known \emph{Team Theory}, which developed out of the need for a mathematical model of cooperating teams within an organization in which all team members have the same object but with different information \cite{han2012game}.
In principle, based on \emph{Team Theory}, the cooperation among neighboring devices can improve the overall performance in automation and economics areas. Interestingly, \emph{Team Theory} can also be used in wireless communications to help the devices select the most efficient parameters, such as power control, beamforming design and time-frequency resource utilization.
For instance, the problem of cooperation in the multicell MIMO downlink precoding under distributed CSI was formulated as a team decision problem in \cite{5454132}.
In addition, utilizing \emph{Team Theory} to model robust coordination, the authors of \cite{gesbert2018team} designed the decentralized MIMO precoding in wireless networks.
Very recently, the authors of \cite{miretti2021team} derived the optimal downlink precoding of CF massive MIMO systems based on the \emph{Team Theory}.
To the best of our knowledge, however, no analysis has been done for the \emph{uplink} combiner design of CF massive MIMO systems with the \emph{Team Theory}.

Motivated by the aforementioned analysis, we utilize the \emph{Team Theory} to design a general uplink combiner framework of CF massive MIMO systems for jointly covering limited data and flexible CSI sharing. Considering both channel estimation errors and pilot contamination, we derive the TMMSE combining of CF massive MIMO systems under unidirectional, centralized and statistical CSI sharing schemes, respectively. Our results show that the performance of centralized TMMSE outperforms the one of unidirectional TMMSE, and the performance gap between them can be efficiently reduced by increasing the number of antennas at the APs.

\textit{Notation:} We use boldface lowercase letters $\mathbf{x}$ and boldface uppercase letters $\mathbf{X}$ to represent column vectors and matrices, respectively.
The $n\times n$ identity matrix is ${{\mathbf{I}}_n}$, the $k$th column of ${{\mathbf{I}}_K}$ is ${{\mathbf{e}}_k}$.
Superscripts $x^\mathrm{*}$, $\mathbf{x}^\mathrm{T}$ and $\mathbf{x}^\mathrm{H}$ are used to denote conjugate, transpose and conjugate transpose, respectively.
The absolute value, the Euclidean norm, the trace operator and the definitions are denoted by $\left|  \cdot  \right|$, $\left\|  \cdot  \right\|$, ${\text{tr}}\left(  \cdot  \right)$, and $\triangleq$, respectively.
Finally, $x \sim \mathcal{C}\mathcal{N}\left( {0,{\sigma^2}} \right)$ represents a circularly symmetric complex Gaussian random variable $x$ with variance $\sigma^2$.


\section{System Model}\label{se:model}

\begin{figure}[t]
\centering
\includegraphics[scale=0.55]{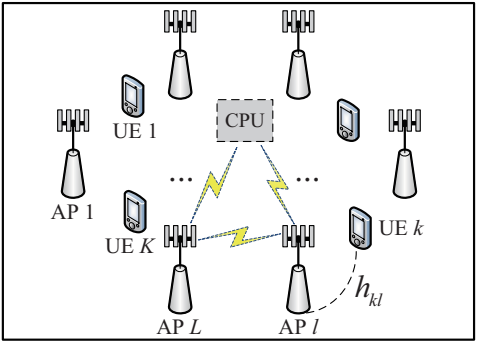}
\caption{Cooperation among APs in CF massive MIMO systems.} \vspace{-4mm}
\label{unidirectional}
\end{figure}

We consider a CF massive MIMO system, where all $L$ geographically distributed APs simultaneously serve all $K$ UEs on the same time-frequency resource. We assume each AP is equipped with $N$ antennas, and each UE is equipped with single antennas.
As illustrated in Fig.~\ref{unidirectional}, the APs can be directly connected to CPU via fronthaul links\footnote{The fronthaul data delivery among the CPU and APs can be carried over wireless links, which is more cost-effective and flexible than the conventional wired fronthaul solutions \cite{8283646}.}, which is a widely used connection scheme in CF massive MIMO systems. Moreover, we use the standard block fading model, where the uplink channel estimation phase occupies $\tau_p$ channel uses and the uplink data transmission phase occupies $\tau_c - \tau_p$ channel uses. In each block, we model the channel between AP $l$ and UE $k$ as the Rayleigh fading:
\begin{align}
{{\mathbf{h}}_{kl}} \sim \mathcal{C}\mathcal{N}\left( {{\mathbf{0}},{{\mathbf{R}}_{kl}}} \right),
\end{align}
where ${{\mathbf{R}}_{kl}} \in {\mathbb{C}^{N \times N}}$ is the spatial correlation matrix, and ${\beta _{kl}} \triangleq {\text{tr}}\left( {{{\mathbf{R}}_{kl}}} \right)/N$ is the large-scale fading coefficient.

\subsection{Uplink Channel Estimation}

We assume that $\tau_p$ mutually orthogonal time-multiplexed pilot sequence ${\phi _1}, \ldots ,{\phi _{{\tau _p}}}$ with ${\left\| {{\phi _t}} \right\|^2} = {\tau _p}$ are utilized, and a large network with $K > \tau_p$ is considered so that the same pilot sequence is used by different UEs. Let ${t_k} \in \left\{ {1, \ldots ,{\tau _p}} \right\}$ denote the index of the pilot used by UE $k$, and the other UEs assigned to the same pilot as UE $k$ is denoted by ${\mathcal{P}_k} = \{ i : t_i = t_k \} \subset \left\{ {1, \ldots ,K} \right\}$.
Then, we can obtain the received signal of AP $l$ from all UEs as
\begin{align}
{{\mathbf{Z}}_l} = \sum\limits_{i = 1}^K {\sqrt p {{\mathbf{h}}_{il}}\phi _{{t_i}}^{\text{T}}}  + {{\mathbf{N}}_l},
\end{align}
where $p>0$ is the transmit power, ${{\bf{N}}_l} \in {\mathbb{C}^{N \times {\tau _p}}}$ is the receiver noise with independent ${\cal C}{\cal N}\left( {{{0}},{\sigma ^2}} \right)$ entries, and $\sigma^2$ is the noise power.
For estimating ${\mathbf{h}}_{kl}$, the received signal at APs is correlated with the associated normalized pilot signal ${\phi _{{t_k}}}/\sqrt {{\tau _p}}$ to obtain
\begin{align}
  {{\mathbf{z}}_{t_kl}} &= \frac{1}{{\sqrt {{\tau _p}} }}{{\mathbf{Z}}_l}\phi _{{t_k}}^* = \sum\limits_{i = 1}^K {\frac{{\sqrt p }}{{\sqrt {{\tau _p}} }}{{\mathbf{h}}_{il}}\phi _{{t_i}}^{\text{T}}} \phi _{{t_k}}^* + \frac{1}{{\sqrt {{\tau _p}} }}{{\mathbf{N}}_l}\phi _{{t_k}}^* \notag \\
   &= \sum\limits_{i \in {\mathcal{P}_k}} {\sqrt {p{\tau _p}} {{\mathbf{h}}_{il}}}  + {{\mathbf{n}}_{{t_k}l}}.
\end{align}
Then, using standard MMSE estimation, the AP $l$ computes the MMSE estimate as
\begin{align}
{{{\mathbf{\hat h}}}_{kl}} = \sqrt {{p}{\tau _p}} {{\mathbf{R}}_{kl}}{{\mathbf{\Psi }}_{kl}}{{\mathbf{z}}_{{t_k}l}},
\end{align}
where
\begin{align}
{{\mathbf{\Psi }}_{kl}} = {\left( {\sum\limits_{i \in {\mathcal{P}_k}} {{p}{\tau _p}{{\mathbf{R}}_{il}}}  + {\sigma ^2}{{\mathbf{I}}_N}} \right)^{ - 1}}.
\end{align}
In addition, the estimate ${{{\mathbf{\hat h}}}_{kl}}$ and estimation errors ${{{\mathbf{\tilde h}}}_{kl}}$ are \mbox{independent} vectors distributed as ${{{\mathbf{\hat h}}}_{kl}} \sim \mathcal{C}\mathcal{N}\left( {{\mathbf{0}},{{\mathbf{Q}}_{kl}}} \right)$ and ${{{\mathbf{\tilde h}}}_{kl}} = \left({{{\mathbf{h}}}_{kl}} - {{{\mathbf{\hat h}}}_{kl}}\right) \sim \mathcal{C}\mathcal{N}\left( {{\mathbf{0}},{{\mathbf{R}}_{kl}} - {{\mathbf{Q}}_{kl}}} \right)$ with
\begin{align}
{{\mathbf{Q}}_{kl}} = {p}{\tau _p}{{\mathbf{R}}_{kl}}{{\mathbf{\Psi }}_{kl}}{{\mathbf{R}}_{kl}}.
\end{align}

\subsection{Uplink Data Transmission and Combining}

Under the uplink data transmission phase, the received complex baseband signal at AP $l$ is
\begin{align}
{{\mathbf{y}}_l} = \sum\limits_{i = 1}^K {{{\mathbf{h}}_{il}}\sqrt p {s_i} + {{\mathbf{n}}_l}},
\end{align}
where $s_i \sim \mathcal{C}\mathcal{N}\left( {0,1} \right)$ is the transmit signal with the data transmit power $p$, and ${{\mathbf{n}}_l}\sim\mathcal{C}\mathcal{N}\left( {{\mathbf{0}},{\sigma ^2}{{\mathbf{I}}_N}} \right)$ is the receiver noise.
To detect the symbol transmitted from the $k$th UE, the received signal ${{\mathbf{y}}_l}$ is multiplied with the conjugate of combining vector. Sending the obtained quantity ${{\overset{\lower0.5em\hbox{$\smash{\scriptscriptstyle\smile}$}}{s} }_{kl}} = {\mathbf{v}}_{kl}^{\text{H}}{{\mathbf{y}}_l}$ to the CPU via the fronthaul, then ${{\hat s}_k}$ is obtained as
\begin{align}\label{shat}
  {{\hat s}_k} &= \frac{1}{{\sqrt {{p}} }}\sum\limits_{l = 1}^L {{\mathbf{v}}_{kl}^{\text{H}}{{\mathbf{y}}_l}} \notag \\
   &= \frac{1}{{\sqrt {{p}} }}\sum\limits_{l = 1}^L {\sum\limits_{i = 1}^K {\sqrt {{p}} {\mathbf{v}}_{kl}^{\text{H}}{{\mathbf{h}}_{il}}{s_i}} }  + \frac{1}{{\sqrt {{p}} }}\sum\limits_{l = 1}^L {{\mathbf{v}}_{kl}^{\text{H}}{{\mathbf{n}}_l}} .
\end{align}
Based on the obtained CSI with estimation errors and pilot contamination at the $l$th AP, we utilize the \emph{Team Theory} to design the TMMSE combining in the following.

\section{Performance Analysis}\label{se:performance}

We use the standard information inequalities to obtain \cite{el2011network}:
\begin{align}
I\left( {{s_k};{{\hat s}_k}} \right) \geqslant  - \log \left( {\mathbb{E}\left\{ {{{\left| {{s_k} - \alpha {{\hat s}_k}} \right|}^2}} \right\}} \right).
\end{align}
Choosing $\alpha  = {\alpha ^*}$ with ${\alpha ^ \star } = \mathbb{E}\left\{ {{s_k}\hat s_k^*} \right\}/\mathbb{E}\left\{ {{{\left| {{{\hat s}_k}} \right|}^2}} \right\}$ being the solution of ${\min _\alpha }\mathbb{E}\left\{ {{{\left| {{s_k} - \alpha {{\hat s}_k}} \right|}^2}} \right\}$, leads to the well-known UatF bound as
\begin{align}
{\text{S}}{{\text{E}}_k^{{\text{UatF}}}} =  - \frac{\tau_c - \tau_p}{\tau_c} \log \left( {\mathbb{E}\left\{ {{{\left| {{s_k} - {\alpha ^*}{{\hat s}_k}} \right|}^2}} \right\}} \right) .
\end{align}
\begin{rem}\label{alpha}
Because $\alpha ^*$ is a scalar that depends only on channel statistics, we have that if ${\mathbf{v}}_{kl}$ is the optimal solution of the following problem:
\begin{align}\label{MSE}
{\mathrm{minimize}}\;{\mathrm{MS}}{{\mathrm{E}}_k}\left( {{\mathbf{v}_{kl}}} \right) = {\mathbb{E}\left\{ {{{\left| {{s_k} - {{\hat s}_k} \left( {\mathbf{v}_{kl}} \right)} \right|}^2}} \right\}} ,
\end{align}
it is also an optimal solution to the problem:
\begin{align}
\mathrm{maximize}\;{\mathrm{S}}{{\mathrm{E}}_k^{{\mathrm{UatF}}}}\left( {{{\mathbf{v}}_{kl}}} \right).
\end{align}
\end{rem}
\begin{thm}
Based on Remark \ref{alpha}, we study the following novel MMSE combining design criterion \cite{9238440}:
\begin{align}\label{minmize}
&{\mathrm{minimize}}\;{\mathrm{MS}}{{\mathrm{E}}_k}\left( {{{\mathbf{v}}_{kl}}} \right) = \mathbb{E}\left\{ {{{\left| {\sum\limits_{l = 1}^L {{\mathbf{v}}_{kl}^{\mathrm{H}}{{\mathbf{h}}_{kl}}}  - 1} \right|}^2}} \right\} \notag\\
&+ \sum\limits_{i \ne k}^K {\mathbb{E}\left\{ {{{\left| {\sum\limits_{l = 1}^L {{\mathbf{v}}_{kl}^{\mathrm{H}}{{\mathbf{h}}_{il}}} } \right|}^2}} \right\}}  + \frac{{{\sigma ^2}}}{p}\sum\limits_{l = 1}^L {\mathbb{E}\left\{ {{{\left\| {{\mathbf{v}}_{kl}^{\mathrm{H}}} \right\|}^2}} \right\}},
\end{align}
where ${{\mathbf{v}}_{kl}} = {f_{kl}}\left( {{{{\mathbf{\hat H}}}_l}} \right)$, with
\begin{align}
{{{\mathbf{\hat H}}}_l} = \left[ {{{{\mathbf{\hat H}}}_{l,1}}, \ldots ,{{{\mathbf{\hat H}}}_{l,L}}} \right], l = 1, \ldots ,L .
\end{align}
Besides, ${{{\mathbf{\hat H}}}_{l,j}} \!=\! {\left[ {{{{\mathbf{\hat h}}}_{1j}}, \ldots ,{{{\mathbf{\hat h}}}_{Kj}}} \right]^{\mathrm{T}}} \!\in\! {\mathbb{C}^{K \times N}}$ denotes the channel estimate of $j$-th AP available at AP $l$.
The corresponding channel gain is represented as ${{\mathbf{H}}_j} \!\!=\!\! {\left[ {{{\mathbf{h}}_{1j}}, \ldots ,{{\mathbf{h}}_{Kj}}} \right]^{\mathrm{T}}} \!\in\! {\mathbb{C}^{K \times N}}$.
\end{thm}
\begin{IEEEproof}
Please refer to Appendix A.
\end{IEEEproof}

Problem \eqref{minmize} is the known family of team decision problems, which are generally difficult to solve for general information constraints. However, we recognize that Problem \eqref{minmize} belongs to the class of quadratic teams. This class exhibits strong structural properties, in particular related to the following solution concept:

The combining function ${f_{kl}^*}$ is a stationary solution for Problem \eqref{minmize} if ${{\text{MSE}}_k}\left( {{f_{kl}^*}} \right) < \infty$ and if the following set of equalities hold
\begin{align}\label{condition}
{\nabla _{f_{kl}^*\left( {{{{\mathbf{\hat H}}}_l}} \right)}}\mathbb{E}\left\{ {{\text{MS}}{{\text{E}}_k}\left| {{{{\mathbf{\hat H}}}_l}} \right.} \right\} = 0,l = 1, \ldots ,L .
\end{align}
Submitting \eqref{minmize} into \eqref{condition}, the conditions in \eqref{condition} can be evaluated as \eqref{ss} at the top of next page.
\newcounter{mytempeqncnt}
\begin{figure*}[t!]
\normalsize
\setcounter{mytempeqncnt}{0}
\setcounter{equation}{15}
\begin{align}\label{ss}
 - \mathbb{E}\left\{ {{{\mathbf{h}}_{kl}}\left| {{{{\mathbf{\hat H}}}_l}} \right.} \right\} \!+\! \sum\limits_{i = 1}^K {\mathbb{E}\left\{ {{{\mathbf{h}}_{il}}{\mathbf{h}}_{il}^{\text{H}}\left| {{{{\mathbf{\hat H}}}_l}} \right.} \right\}} f_{kl}^*\left( {{{{\mathbf{\hat H}}}_l}} \right) \!+\! \sum\limits_{j \ne l}^L {\sum\limits_{i = 1}^K {\mathbb{E}\left\{ {{{\mathbf{h}}_{il}}{\mathbf{h}}_{ij}^{\text{H}}f_{kj}^*\left( {{{{\mathbf{\hat H}}}_j}} \right)\left| {{{{\mathbf{\hat H}}}_l}} \right.} \right\}} }  \!+\! \frac{{{\sigma ^2}}}{p}{{\mathbf{I}}_N}f_{kl}^*\left( {{{{\mathbf{\hat H}}}_l}} \right) = {\mathbf{0}},l = 1, \ldots ,L ,
\end{align}
\setcounter{equation}{16}
\hrulefill
\end{figure*}
Making use of the law of total expectation and the available CSI, we then compute the third term of \eqref{ss} as
\begin{align}
  &\mathbb{E}\left\{ {{{\mathbf{h}}_{il}}{\mathbf{h}}_{ij}^{\text{H}}f_{kj}^*\left( {{{{\mathbf{\hat H}}}_j}} \right)\left| {{{{\mathbf{\hat H}}}_l}} \right.} \right\} \notag\\
   &= \mathbb{E}\left\{ {\mathbb{E}\left\{ {{{\mathbf{h}}_{il}}{\mathbf{h}}_{ij}^{\text{H}}\left| {{{{\mathbf{\hat h}}}_{il}},{{{\mathbf{\hat h}}}_{ij}}} \right.} \right\}f_{kj}^*\left( {{{{\mathbf{\hat H}}}_j}} \right)\left| {{{{\mathbf{\hat H}}}_l}} \right.} \right\} \notag \\
   &= \mathbb{E}\left\{ {\mathbb{E}\left\{ {{{\mathbf{h}}_{il}}\left| {{{{\mathbf{\hat h}}}_{il}},{{{\mathbf{\hat h}}}_{ij}}} \right.} \right\}\mathbb{E}\left\{ {{\mathbf{h}}_{ij}^{\text{H}}\left| {{{{\mathbf{\hat h}}}_{il}},{{{\mathbf{\hat h}}}_{ij}}} \right.} \right\}f_{kj}^*\left( {{{{\mathbf{\hat H}}}_j}} \right)\left| {{{{\mathbf{\hat H}}}_l}} \right.} \right\} \notag \\
   &= \mathbb{E}\left\{ {\mathbb{E}\left\{ {{{\mathbf{h}}_{il}}\left| {{{{\mathbf{\hat h}}}_{il}}} \right.} \right\}\mathbb{E}\left\{ {{\mathbf{h}}_{ij}^{\text{H}}\left| {{{{\mathbf{\hat h}}}_{ij}}} \right.} \right\}f_{kj}^*\left( {{{{\mathbf{\hat H}}}_j}} \right)\left| {{{{\mathbf{\hat H}}}_l}} \right.} \right\} \notag \\
   &= {{{\mathbf{\hat h}}}_{il}}\mathbb{E}\left\{ {{\mathbf{\hat h}}_{ij}^{\text{H}}f_{kj}^*\left( {{{{\mathbf{\hat H}}}_j}} \right)\left| {{{{\mathbf{\hat H}}}_l}} \right.} \right\} .
\end{align}
Due to ${{{\mathbf{\hat h}}}_{il}}$ and ${{{\mathbf{\tilde h}}}_{il}}$ are independent in MMSE estimation, we compute the second term of \eqref{ss} as
\begin{align}
  &\mathbb{E}\left\{ {{{\mathbf{h}}_{il}}{\mathbf{h}}_{il}^{\text{H}}\left| {{{{\mathbf{\hat H}}}_l}} \right.} \right\} = \mathbb{E}\left\{ {{{\mathbf{h}}_{il}}{\mathbf{h}}_{il}^{\text{H}}\left| {{{{\mathbf{\hat H}}}_l}} \right.} \right\} \notag \\
   &= \mathbb{E}\left\{ {\left( {{{{\mathbf{\hat h}}}_{il}} + {{{\mathbf{\tilde h}}}_{il}}} \right){{\left( {{{{\mathbf{\hat h}}}_{il}} + {{{\mathbf{\tilde h}}}_{il}}} \right)}^{\text{H}}}\left| {{{{\mathbf{\hat H}}}_l}} \right.} \right\} \notag \\
   &= {{{\mathbf{\hat h}}}_{il}}{\mathbf{\hat h}}_{il}^{\text{H}}{\text{ + }}{{\mathbf{C}}_{il}} .
\end{align}
We also have $\mathbb{E}\left\{ {{{\mathbf{h}}_{kl}}\left| {{{{\mathbf{\hat H}}}_l}} \right.} \right\} = {{{\mathbf{\hat h}}}_{kl}}$.
Furthermore, we can derive the combining functions in the form of set of equalities as
\begin{align}\label{f_H}
f_{kl}^*\left( {{{{\mathbf{\hat H}}}_l}} \right) \!=\! {{\mathbf{A}}_l}{{\left(\! {{{\mathbf{e}}_k} \!-\! \sum\limits_{j \ne l}^L {\mathbb{E}\left\{ {{\mathbf{\hat H}}_{j,j}f_{kj}^*\left( {{{{\mathbf{\hat H}}}_j}} \right)\left| {{{{\mathbf{\hat H}}}_l}} \right.} \right\}} } \!\right)}}, \forall l ,
\end{align}
where
\begin{align}
{{\mathbf{A}}_l} = {\left( {\sum\limits_{i = 1}^K {\left( {{{\mathbf{\hat h}}}_{il}}{\mathbf{\hat h}}_{il}^{\text{H}}{\text{  +  }}{{\mathbf{C}}_{il}} \right) + \frac{{{\sigma ^2}}}{p}{{\mathbf{I}}_N} } } \right)^{ - 1}}{{{\mathbf{\hat H}}}_{l,l}^\text{H}} .
\end{align}
\begin{rem}\label{rem2}
Equation \eqref{f_H} can be divided into two parts. The matrix ${{\mathbf{A}}_l}$ can be recognized as a local MMSE combining stage, and the rest can be then interpreted as a corrective stage which takes into account the effect of the other APs based on the available CSI and long-term statistical information.
\end{rem}

\subsection{Unidirectional CSI Sharing}

We first consider the local channel measurements to be shared unidirectionally along a serial fronthaul, which also is known as a radio stripe \cite{9499049}. Specifically, this particular information structure can be expressed as
\begin{align}\label{UIS}
{{{\mathbf{\hat H}}}_{l,j}} = \left\{ {\begin{array}{*{20}{c}}
  {{{{\mathbf{\hat H}}}_{j,j}},j \leqslant l} \\
  {\mathbb{E}\left\{ {{{\mathbf{H}}_j}} \right\},j > l}
\end{array}} \right. .
\end{align}
\begin{thm}\label{thm2}
The unidirectional TMMSE combining solving \eqref{f_H} under the unidirectional CSI sharing \eqref{UIS} is given by
\begin{align}\label{f_uni}
f_{kl}^*\left( {{{{\mathbf{\hat H}}}_l}} \right) = {{\mathbf{A}}_l}{{\mathbf{S}}_l}\prod\limits_{s = 1}^{l - 1} {{{{\mathbf{\bar S}}}_s}} {{\mathbf{e}}_k} ,
\end{align}
where
\begin{align}
  {{{\mathbf{\bar S}}}_l} &= {{\mathbf{I}}_K} - {{\mathbf{\Lambda }}_l}{{\mathbf{S}}_l}, \hfill \\
  {{\mathbf{S}}_l} &= {\left( {{{\mathbf{I}}_K} - {{\mathbf{\Pi }}_l}{{\mathbf{\Lambda }}_l}} \right)^{ - 1}}\left( {{{\mathbf{I}}_K} - {{\mathbf{\Pi }}_l}} \right), \hfill \\
  {{\mathbf{\Pi }}_l} &= \mathbb{E}\left\{ {{{\mathbf{\Lambda }}_{l+1}}{{\mathbf{S}}_{l + 1}}} \right\} \!+\! {{\mathbf{\Pi }}_{l + 1}}\mathbb{E}\left\{ {{{{\mathbf{\bar S}}}_{l + 1}}} \right\},{{\mathbf{\Pi }}_L} \!=\! {\mathbf{0}} , \\
  {{\mathbf{\Lambda }}_l} &= {\mathbf{\hat H}}_{l,l}{{\mathbf{A}}_l} .
\end{align}
\end{thm}
\begin{IEEEproof}
Please refer to Appendix B.
\end{IEEEproof}

\begin{cor}
Under the unidirectional CSI sharing scheme in Theorem \ref{thm2}, we rearrange the order of APs according to the sorting rule: If $\sum\nolimits_{k = 1}^K {{\beta _{kl}}}  > \sum\nolimits_{k = 1}^K {{\beta _{kj}}} $, we set the CSI sharing order of AP $l$ and AP $j$ as $l<j$.
That is, the AP with poor CSI needs to know more CSI of other APs.
\end{cor}

\subsection{Centralized CSI Sharing}

For the centralized CSI sharing scheme, each AP knows the CSI of all other APs. Then, we have
\begin{align}\label{centralized}
{{{\mathbf{\hat H}}}_{l,j}} = {{{\mathbf{\hat H}}}_{j,j}},\forall j .
\end{align}
Then, the centralized TMMSE combining solving \eqref{f_H} under the centralized CSI sharing \eqref{centralized} is given by
\begin{align}\label{f_centralized}
f_{kl}^*\left( {{{{\mathbf{\hat H}}}_l}} \right) = {{\mathbf{A}}_l}{{\mathbf{a}}_{kl}} .
\end{align}
Submitting \eqref{f_centralized} into \eqref{f_H}, we can obtain
\begin{align}\label{c_equations}
{{\mathbf{a}}_{kl}} + \sum\limits_{j \ne l}^L {\mathbb{E}\left\{ {{{\mathbf{\Lambda }}_j}{{\mathbf{a}}_{kj}}\left| {{{{\mathbf{\hat H}}}_l}} \right.} \right\}}  = {{\mathbf{e}}_k},l = 1, \ldots ,L ,
\end{align}
With the help of the centralized CSI sharing \eqref{centralized}, equations \eqref{c_equations} can be written as
\begin{align}
{{\mathbf{a}}_{kl}} + \sum\limits_{j \ne l}^L {{{\mathbf{\Lambda }}_j}{{\mathbf{a}}_{kj}}}  = {{\mathbf{e}}_k},l = 1, \ldots ,L ,
\end{align}
Finally, we can derive ${{{\mathbf{a}}_{kl}}}$ as
\begin{align}
\left[\! {\begin{array}{*{20}{c}}
  {{{\mathbf{a}}_{k1}}} \\
   \vdots  \\
  {{{\mathbf{a}}_{k\left( {L - 1} \right)}}} \\
  {{{\mathbf{a}}_{kL}}}
\end{array}} \!\right] \!=\! {\left[\! {\begin{array}{*{20}{c}}
  {{{\mathbf{I}}_K}}& \ldots &{{{\mathbf{\Lambda }}_{L - 1}}}&{{{\mathbf{\Lambda }}_L}} \\
   \vdots & \ddots & \vdots & \vdots  \\
  {{{\mathbf{\Lambda }}_1}}& \ldots &{{{\mathbf{I}}_K}}&{{{\mathbf{\Lambda }}_L}} \\
  {{{\mathbf{\Lambda }}_1}}& \ldots &{{{\mathbf{\Lambda }}_{L - 1}}}&{{{\mathbf{I}}_K}}
\end{array}} \!\right]^{ - 1}}\left[\! {\begin{array}{*{20}{c}}
  {{{\mathbf{e}}_k}} \\
   \vdots  \\
  {{{\mathbf{e}}_k}} \\
  {{{\mathbf{e}}_k}}
\end{array}} \!\right] .
\end{align}

\subsection{Statistical CSI Sharing}

For the statistical CSI sharing, each AP knows its local instantaneous CSI and statistical CSI of other APs. This is expressed by formula as
\begin{align}\label{local}
{{{\mathbf{\hat H}}}_{l,j}} = \mathbb{E}\left\{ {{{\mathbf{H}}_j}} \right\},j \ne l .
\end{align}
Therefore, the statistical TMMSE combining solving \eqref{f_H} under the statistical CSI sharing \eqref{local} also can be is expressed as \eqref{f_centralized}. It is similar to the steps of the centralized CSI sharing scheme, we obtain the following equations:
\begin{align}
{{\mathbf{a}}_{kl}} + \sum\limits_{j \ne l}^L {\mathbb{E}\left\{ {{{\mathbf{\Lambda }}_j}} \right\}{{\mathbf{a}}_{kj}}}  = {{\mathbf{e}}_k},l = 1, \ldots ,L ,
\end{align}
which can be used to derive ${\mathbf{a}}_{kl}$.

\section{Numerical Results and Discussion}\label{se:numerical}

Let $L$ APs and $K$ UEs uniformly and independently distribute within a square of size $500 \ {\mathrm{m}}\times 500\ {\mathrm{m}}$ in our simulation setup. Moreover, we make use of the three-slope propagation model from \cite{Ngo2017Cell} as
\begin{align}
{{\beta _{kl}}}\left[ {{\mathrm{dB}}} \right] \!= \!\left\{ {\begin{array}{*{20}{c}}
  { - 81.2,{d_{kl}} < 10{\mathrm{m}}} \\
  { \!\!\!- 61.2 \!- \!20{{\log }_{10}}\!\left( {\frac{{{d_{kl}}}}{{1{\mathrm{m}}}}} \right),10{\mathrm{m}} \!\leqslant {d_{kl}} \!<\! 50{\mathrm{m}}} \\
  { \!- 35.7 \!- \!35{{\log }_{10}}\left( {\frac{{{d_{kl}}}}{{1{\mathrm{m}}}}} \right) \!+ \!{F_{kl}},{d_{kl}} \!\geqslant\! 50{\mathrm{m}}},
\end{array}} \right.
\end{align}
where $d_{kl}$ denotes the horizontal distance between AP $l$ and UE $k$. When the distance is larger than 50m and the shadowing terms ${F_{kl}} \sim \mathcal{N}\left( {0,{8^2}} \right)$ are correlated as
\begin{align}
\mathbb{E}\left\{ {{F_{kl}}{F_{ij}}} \right\} = \frac{{{8^2}}}{2}\left( {{2^{ - {\delta _{ki}}/100{\mathrm{m}}}} + {2^{ - {\upsilon _{lj}}/100{\mathrm{m}}}}} \right),
\end{align}
where $\delta _{ki}$ represents the distance between the $k$th UE and $i$th UE, $\upsilon  _{lj}$ represents the distance between the $l$th AP and $j$th AP.
We assume the carrier frequency is $f_c\!=\!2$ GHz, and the bandwidth is $B\!=\!20$ MHz. Moreover, both the pilot transmit power and the data transmission power are $p \!=\!23$ dBm. Besides, the noise power is $\sigma^2\!=\!-96$ dBm, the coherence block contains $\tau_c \!=\! 200$ channel uses.

\begin{figure}[t]
\centering
\includegraphics[scale=0.5]{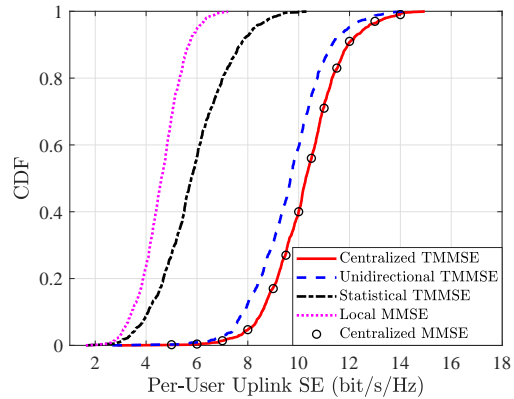}
\caption{CDF of uplink SE for CF massive MIMO systems with TMMSE combining under different CSI sharing schemes ($L=100$, $K=10$, $N=2$, $\tau_p = 10$).} \vspace{-4mm}
\label{figure1}
\end{figure}

\begin{figure}[t]
\centering
\includegraphics[scale=0.5]{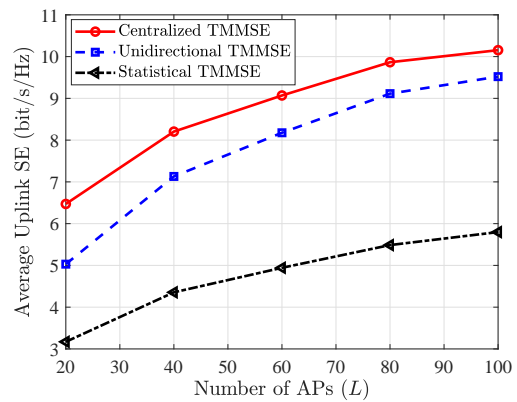}
\caption{Average uplink SE for CF massive MIMO systems against different numbers of APs ($K=10$, $N=2$, $\tau_p=10$).} \vspace{-4mm}
\label{figure2}
\end{figure}

\begin{figure}[t]
\centering
\includegraphics[scale=0.5]{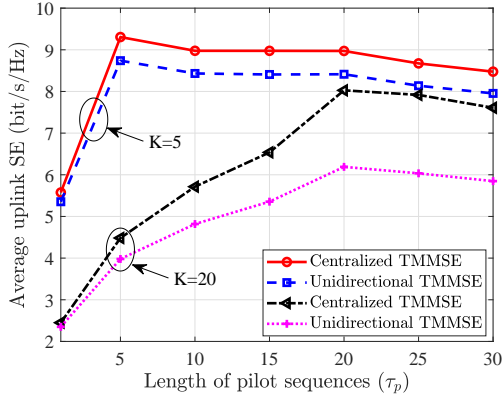}
\caption{Average uplink SE for CF massive MIMO systems against the length of pilot sequence ($L=50$, $N=2$).} \vspace{-4mm}
\label{figure3}
\end{figure}

\begin{figure}[t]
\centering
\includegraphics[scale=0.5]{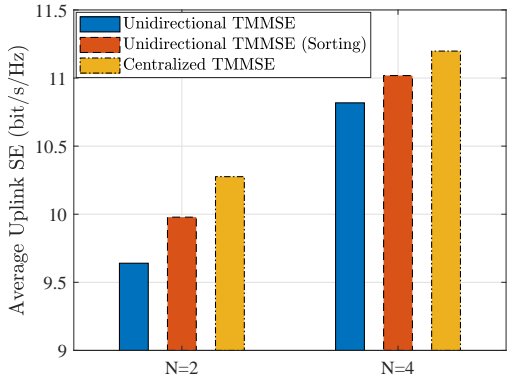}
\caption{Average uplink SE for CF massive MIMO systems with different numbers of antennas per AP ($L=100$, $K=10$, $\tau_p=10$).} \vspace{-4mm}
\label{figure4}
\end{figure}

Fig.~\ref{figure1} compares the CDF of SE for unidirectional, centralized and statistical TMMSE combining schemes, where CF massive MIMO systems operate in unidirectional, centralized and statistical CSI sharing schemes, respectively. In addition, the traditional centralized MMSE and local MMSE combining in \cite{8845768} are considered for comparison. It is found that, compared with centralized TMMSE combining, unidirectional TMMSE combining only have 5.6\% median SE performance loss, but it can save nearly half the cost of CSI sharing. Moreover, we find that statistical TMMSE combining achieves larger uplink SE than local MMSE combining. The reason is that both local estimate and global statistical CSI are used to design statistical TMMSE combiner. Besides, the centralized MMSE combining in \cite{8845768} has the same SE performance as our derived centralized TMMSE combining.

The average uplink SE for CF massive MIMO systems under three different CSI sharing schemes is shown in Fig.~\ref{figure2}, as an increasing function of the number of APs. We notice that the performance gap between unidirectional TMMSE and centralized TMMSE gradually decreases with the increase of the number of APs. For instance, at $L=20$ and $L=100$, the SE gaps are 1.4 bit/s/Hz and 0.6 bit/s/Hz, respectively. However, under the same parameters, the SE gaps between statistical and centralized TMMSE are 3.3 bit/s/HZ and 4.4 bit/s/Hz, respectively.

Fig.~\ref{figure3} illustrates the average uplink SE for CF massive MIMO systems against the length of the pilot sequence. It is clear that the average SE performance of both centralized and unidirectional TMMSE combining increases first and then decreases with the increase of the length of the pilot sequence. The reason is that, while increasing the length of the pilot sequence to reduce pilot contamination, the channel uses for transmitting data are also decreasing. Moreover, the optimal length of the pilot sequence at the point where we achieve maximum SE increases with the increase of the number of UEs. Furthermore, we also find more UEs lead to a larger performance gap between unidirectional TMMSE and centralized TMMSE combining.

Fig.~\ref{figure4} shows the average uplink SE for CF massive MIMO systems with different numbers of antennas per AP. It is clear that the unidirectional TMMSE based on large-scale fading sorting outperforms the original unidirectional TMMSE. The reason is that the weak coverage APs can use more information for designing combiner to reduce the interference from other APs. In addition, we find that increasing the number of antennas at the APs also can reduce the SE gap between unidirectional TMMSE and centralized TMMSE.

\section{Conclusion}\label{se:conclusion}

In this paper, we design the uplink TMMSE combining of CF massive MIMO systems based on the so-called theory of teams. Taking into account both channel estimation errors and pilot contamination, we derive unidirectional, centralized and statistical TMMSE combining schemes, where CF massive MIMO systems operate in unidirectional, centralized and statistical CSI sharing schemes, respectively. It is interesting to find that, compared to centralized TMMSE combining, the unidirectional TMMES combining saves nearly half the cost of CSI sharing while with only limited performance loss. Moreover, the performance gap between unidirectional and centralized TMMSE combiners can be further reduced by increasing the number of APs and antennas per AP.

\vspace{-0.3cm}
\begin{appendices}
\section{Proof of Theorem 1}

Submitting \eqref{shat} into \eqref{MSE}, we derive $\mathbb{E}\left\{ {{{\left| {{s_k} - {{\hat s}_k}} \right|}^2}} \right\}$ as
\begin{align}
\mathbb{E}\left\{ {{{\left| {{s_k} - \sum\limits_{l = 1}^L {\sum\limits_{i = 1}^K {{\mathbf{v}}_{kl}^{\text{H}}{{\mathbf{h}}_{il}}{s_i}} }  - \frac{1}{{\sqrt p }}\sum\limits_{l = 1}^L {{\mathbf{v}}_{kl}^{\text{H}}{{\mathbf{n}}_l}} } \right|}^2}} \right\}. \notag
\end{align}
Due to ${s_i} \sim \mathcal{C}\mathcal{N}\left( {0,1} \right),i = 1, \ldots ,K$ are independent and identically distributed. We further obtain
\begin{align}\label{skshat}
\mathbb{E}&\left\{ {{{\left| {{s_k} - {{\hat s}_k}} \right|}^2}} \right\} = 1 - 2\mathbb{E}\left\{ {\sum\limits_{l = 1}^L {{\mathbf{v}}_{kl}^{\text{H}}{{\mathbf{h}}_{kl}}} } \right\} \notag\\
&+ \sum\limits_{i = 1}^K \mathbb{E} \left\{ {{{\left| {\sum\limits_{l = 1}^L {{\mathbf{v}}_{kl}^{\text{H}}{{\mathbf{h}}_{il}}} } \right|}^2}} \right\} + \frac{{{\sigma ^2}}}{p}\sum\limits_{l = 1}^L {\mathbb{E}\left\{ {{{\left\| {{\mathbf{v}}_{kl}^{\text{H}}} \right\|}^2}} \right\}} .
\end{align}
With the help of the perfect square trinomial, \eqref{skshat} becomes \eqref{minmize} to finish the proof.
In addition, it is worth noting that we also derive
\begin{align}
  \!\mathbb{E}\left\{ {{s_k}\hat s_k^*} \right\} &= \mathbb{E}\left\{ {\sum\limits_{l = 1}^L {{\mathbf{v}}_{kl}^{\text{H}}{{\mathbf{h}}_{kl}}} } \right\} , \hfill \\
 \! \mathbb{E}\!\left\{\! {{{\left| {{{\hat s}_k}} \right|}^2}} \!\right\} &\!=\! \sum\limits_{i = 1}^K \mathbb{E}\! \left\{\! {{{\left| {\sum\limits_{l = 1}^L {{\mathbf{v}}_{kl}^{\text{H}}{{\mathbf{h}}_{il}}} } \right|}^2}} \!\right\} \!+\! \frac{{{\sigma ^2}}}{p}\!\sum\limits_{l = 1}^L \!{\mathbb{E}\!\left\{ {{{\left\| {{\mathbf{v}}_{kl}^{\text{H}}} \right\|}^2}} \right\}} ,
\end{align}
to calculate the value of $\alpha^*$.
\section{Proof of Theorem 2}

Based on the Remark \ref{rem2} and the unidirectional CSI sharing scheme \eqref{UIS}, where the AP $l$ only knows the CSI of AP $j$ ($j \leqslant l$). Therefore, we assume the unidirectional TMMSE combining at AP $l$ is in the form of \eqref{f_uni}. Then, by submitting \eqref{f_uni} into \eqref{f_H}, we can derive
\begin{align}\label{IK}
&\left( {{{\mathbf{S}}_l} + \sum\limits_{j > l}^L {\mathbb{E}\left\{ {{{\mathbf{\Lambda }}_j}{{\mathbf{S}}_j}\prod\limits_{s = l + 1}^{j - 1} {{{{\mathbf{\bar S}}}_s}} } \right\}{{{\mathbf{\bar S}}}_l}} } \right)\prod\limits_{s = 1}^{l - 1} {{{{\mathbf{\bar S}}}_s}} \notag\\
 &+ \sum\limits_{j < l}^L {{{\mathbf{\Lambda }}_j}{{\mathbf{S}}_j}} \prod\limits_{s = 1}^{j - 1} {{{{\mathbf{\bar S}}}_s}}  = {{\mathbf{I}}_K} .
\end{align}
Since ${{{\mathbf{S}}_i}}$ and ${{{{\mathbf{\bar S}}}_i}}$ are independent from ${{{\mathbf{S}}_j}}$ and ${{{{\mathbf{\bar S}}}_j}}$ for $i \ne j$, we have
\begin{align}
  &\sum\limits_{j > l}^L {\mathbb{E}\left\{ {{{\mathbf{\Lambda }}_j}{{\mathbf{S}}_j}\prod\limits_{s = l + 1}^{j - 1} {{{{\mathbf{\bar S}}}_s}} } \right\}} = \sum\limits_{j > l}^L {\mathbb{E}\left\{ {{{\mathbf{\Lambda }}_j}{{\mathbf{S}}_j}} \right\}\prod\limits_{s = l + 1}^{j - 1} {\mathbb{E}\left\{ {{{{\mathbf{\bar S}}}_s}} \right\}} }  \notag \\
  & = \mathbb{E}\left\{ {{{\mathbf{\Lambda }}_{l+1}}{{\mathbf{S}}_{l + 1}}} \right\} + \sum\limits_{j > l + 1}^L {\mathbb{E}\left\{ {{{\mathbf{\Lambda }}_j}{{\mathbf{S}}_j}} \right\}\prod\limits_{s = l + 1}^{j - 1} {\mathbb{E}\left\{ {{{{\mathbf{\bar S}}}_s}} \right\}} }  \notag \\
   &= \mathbb{E}\left\{ {{{\mathbf{\Lambda }}_{l+1}}{{\mathbf{S}}_{l + 1}}} \right\} \notag\\
   &\;\;\;+ \left( {\sum\limits_{j > l + 1}^L {\mathbb{E}\left\{ {{{\mathbf{\Lambda }}_j}{{\mathbf{S}}_j}} \right\}\prod\limits_{s = l + 2}^{j - 1} {\mathbb{E}\left\{ {{{{\mathbf{\bar S}}}_s}} \right\}} } } \right)\mathbb{E}\left\{ {{{{\mathbf{\bar S}}}_{l + 1}}} \right\} \hfill .
\end{align}
The second and last terms of the above chain of equalities define a recursion terminating with $\mathbb{E}\left\{ {{{\mathbf{\Lambda }}_L}{{\mathbf{S}}_L}} \right\} + {\mathbf{0}}\mathbb{E}\left\{ {{{{\mathbf{\bar S}}}_L}} \right\} = {{\mathbf{\Pi }}_{L - 1}}$. This recursion gives precisely
\begin{align}
\sum\limits_{j > l}^L {\mathbb{E}\left\{ {{{\mathbf{\Lambda }}_j}{{\mathbf{S}}_j}} \right\}\prod\limits_{s = l + 1}^{j - 1} {\mathbb{E}\left\{ {{{{\mathbf{\bar S}}}_s}} \right\}} }  = {{\mathbf{\Pi }}_l} .
\end{align}
With the property ${{\mathbf{S}}_l} \!+\! {{\mathbf{\Pi }}_l}{{{\mathbf{\bar S}}}_l} \!=\! {{\mathbf{I}}_K}$, \eqref{IK} can be simplified to
\begin{align}
  &\prod\limits_{s = 1}^{l - 1} {{{{\mathbf{\bar S}}}_s}}  + \sum\limits_{j < l}^L {{{\mathbf{\Lambda }}_j}{{\mathbf{S}}_j}} \prod\limits_{s = 1}^{j - 1} {{{{\mathbf{\bar S}}}_s}}  = {{{\mathbf{\bar S}}}_{l - 1}}\prod\limits_{s = 1}^{l - 2} {{{{\mathbf{\bar S}}}_s}}  \notag\\
  &+ {{\mathbf{\Lambda }}_{l-1}}{{\mathbf{S}}_{l - 1}}\prod\limits_{s = 1}^{l - 2} {{{{\mathbf{\bar S}}}_s}}  + \sum\limits_{j < l - 1}^L {{{\mathbf{\Lambda }}_j}{{\mathbf{S}}_j}} \prod\limits_{s = 1}^{j - 1} {{{{\mathbf{\bar S}}}_s}}  \notag \\
   &= \left( {{{{\mathbf{\bar S}}}_{l - 1}} + {{\mathbf{\Lambda }}_{l-1}}{{\mathbf{S}}_{l - 1}}} \right) \prod\limits_{s = 1}^{l - 2} {{{{\mathbf{\bar S}}}_s}} + \sum\limits_{j < l - 1}^L {{{\mathbf{\Lambda }}_j}{{\mathbf{S}}_j}} \prod\limits_{s = 1}^{j - 1} {{{{\mathbf{\bar S}}}_s}}  \notag\\
   &= \prod\limits_{s = 1}^{l - 2} {{{{\mathbf{\bar S}}}_s}} + \sum\limits_{j < l - 1}^L {{{\mathbf{\Lambda }}_j}{{\mathbf{S}}_j}} \prod\limits_{s = 1}^{j - 1} {{{{\mathbf{\bar S}}}_s}} ,
\end{align}
where the last equation follows the definition of ${{{{\mathbf{\bar S}}}_l}}$, and we identify another recursive structure among the remaining terms. By continuing until termination, we finally obtain $\prod\limits_{s = 1}^{l - 1} {{{{\mathbf{\bar S}}}_s}}  + \sum\limits_{j < l}^L {{{\mathbf{\Lambda }}_j}{{\mathbf{S}}_j}} \prod\limits_{s = 1}^{j - 1} {{{{\mathbf{\bar S}}}_s}}  = {{\mathbf{I}}_K}$, which proves the key statement under the assumption that all the matrix inverses involved exist.
\end{appendices}

\vspace{0.3cm}
\bibliographystyle{IEEEtran}
\bibliography{IEEEabrv,Ref}

\end{document}